\documentclass{article}    

\usepackage{graphicx}

\title{\textbf{The Cat oRules}}  
\author{Richard Mould\footnote{Department of Physics and Astronomy, State University of New York, Stony Brook,
\mbox{New York} 11794-3800; http://ms.cc.sunysb.edu/\~{}rmould}}  
\date{}    
   
\begin{document}             

\maketitle              

\begin{abstract}

The oRules of state reduction are applied to the Schr\"{o}dinger cat experiment.  It is shown that these rules can unambiguously
describe the conscious state of the cat, as well as an outside observer at any time during the experiment.  Two versions of the
experiment are considered.  In version I, the conscious cat is made unconscious by a mechanism that is triggered by a radioactive
decay. In version II, the sleeping cat is made conscious by an alarm clock that is triggered by a radioactive decay.

\end{abstract}

\section*{Introduction}
Four rules  called the oRules are given in previous papers \cite{RM1, RM2}.  These rules are said to govern the
process of stochastic choice and state reduction in an \emph{ontological model} of a quantum mechanical system, and describe how
the consciousness awareness of an observer changes  during this process.  In the present paper, these rules are applied to
the case of Schr\"{o}dinger's cat. 

An early version of the oRules appears in another paper were they are simply called \emph{The Rules} \cite{RM3}.  This early
version is also applied to the Schr\"{o}dinger cat \cite{RM4}.  The present paper is a new portrayal of ref.\ 4 that uses a better
representation of continuous change, plus the more recent version of the rules, attended by a discussion that brings the text up to
date.    

In ref.\ 1, an interaction is studied that involves a particle passing over a detector with some probability of capture.  A
conscious observer witnesses the detector at various times during the interaction.  It is found that if the observer observes the
detector during a particle capture, a new state of consciousness  accompanies the associated state reduction.  That is, when the
oRules are applied in the cases investigated in ref.\ 1, consciousness is found to switch from one state to another on the occasion
of a stochastic choice.          

There are four separate oRules (1-4).  The first refers to the \emph{probability current} $J$ that flows
into a state.  The current $J$ is defined to be the time rate of change of the square modulus.

\vspace{.4cm}

\noindent
\textbf{oRule (1)}: \emph{For any subsystem of n components in a system having a total square modulus equal to s, the probability
per unit time of a stochastic choice of one of those components at time t is given by $(\Sigma_nJ_n)/s$, where the net probability
current $J_n$ going into the $n^{th}$ component at that time is positive}.

\vspace{.4cm}

The \emph{ready brain state} referred to in the next oRule is defined as one that is \textbf{not conscious}, but is
physiologically capable of becoming conscious if it is stochastically chosen.  The \emph{active} brain state referred to below is
one that is either conscious or ready.  Ready brain states are underlined in this treatment whereas conscious brain states are
not.  This switches the convention adopted in ref.\ 3.  

\vspace{.4cm}

\noindent
\textbf{oRule (2)}: \emph{If an interaction produces new components that are discontinuous with the initial state or with each
other, then all of the active brain states in the new components will be ready brain states.}
\begin{center}
(see ref.\ 1 for an elaboration of ``discontinuous" and ``initial state")
\end{center}

\vspace{.4cm}

\noindent
\textbf{oRule (3)}: \emph{If a component containing ready brain states is stochastically chosen, then those states will become
conscious brain states, and all other components in the superposition will be immediately reduced to zero.}

\begin{center}
(see ref.\ 1 for a discussion of ``immediately")
\end{center}

\vspace{.4cm}

\noindent
\textbf{oRule (4)}: \emph{If a component in a superposition is entangled with a ready brain state, then that component can only
receive probability current.}

\vspace{.4cm}

The purpose of the present paper is to apply these rules to two versions of the Schr\"{o}dinger cat experiment.  Version I is a
somewhat modified formulation of that famous puzzle.  That usually involves a cat being placed on two components of a quantum
mechanical superposition, where it is alive on one component and dead on the other.   This distinction is ambiguous because an
alive cat can be unconscious, in which case it is every bit as inert as a dead cat.  The distinction used here is that the cat is
conscious on one component of the superposition, and unconscious on the other.  In version II, the cat begins in an unconscious
state, and is aroused to a conscious state.

\section*{The Apparatus}
We first look at the apparatus that is used in the Schr\"{o}dinger cat experiment \emph{without} a cat or an external observer
being present.  The oRules applied to the apparatus by itself gives an unaccustomed glimpse of the quantum mechanical behavior of a
macroscopic thing.  This happens because state reduction, under the oRules, can only occur when there is an observer present.  So
we will see what a functioning macroscopic object looks like when it carries out a routine without the benefit of a `collapse' of
the state anywhere along the way.  

The apparatus will consist of a radioactive source and a detector that is denoted by either $d_0$ or $d_1$, where the first means
that the detector has not yet captured the decay particle and the second means that it has.  The detector output will be connected
to a mechanical device that carries out a certain task, such as a hammer falling on a container that then releases an anesthetic
gas.  This device will be denoted by $M(\alpha, t)$, where $\alpha$ indicates the extent to which the task has been completed, and
$t$ is the time.  The component $d_0M(\alpha_0, t)$ indicates that the source has not yet decayed at time $t$ and that the
mechanical device is still in its initial position given by $\alpha_0$.  The component $d_1M(\alpha_1, t)$ indicates that the
decay has already occurred by the time $t$, and that the mechanical device has advanced to a position given by $\alpha_1$.  Let
$i_0$ be an indicator that tells us that $M$ has not completed its task, and $i_1$ tells us that it has.  Then $d_1M(\alpha_1,
t)i_0$ means that the device has not completed its task at time $t$.  When $\alpha = \alpha_f$ we will say that the device $M$ has
fully run its course, so $d_1M(\alpha_f, t)i_1$ means that the source has decayed, and that the mechanical device has completed
its task at time $t$ as indicated by $\alpha_f$ and by the indicator $i_1$.  We also suppose that the source is exposed to the
detector for a time that is limited to the half-life of a single emission.  At that time a clock will shut off the detector, so it
will remain in the state $d_0$ if there has not yet been a particle capture.

The system at $t_0 = 0$ is then $\Phi(t_0) = d_0M(\alpha_0, t_0)i_0$, and in time becomes 
\begin{equation}
\Phi(t \ge t_0)= d_0M(\alpha_0, t)i_0 + d_1M(\alpha,t)i_0 \rightarrow d_1M(\alpha_f, t)i_1
\end{equation}
where all but the first component are zero at $t_0$, and where $a_0 \ge \alpha < \alpha_f$.  The arrow represents a
continuous-classical progression from the second to the third term in eq.\ 1, so these two terms are really a single component
seen at two different times.  A plus sign always represents a discontinuous change (here between $d_0$ and $d_1$), and an arrow
always represents a continuous change.

A spread of alphas at a given time represents a possible uncertainty in the state of the mechanical device at that moment. 
Although the device is macroscopic, there is a quantum mechanical uncertainty as to when it begins its operation.  The function
$M(\alpha, t)$ is therefore a pulse that represents that uncertainty moving along the $\alpha$ axis.  Although the second and
third terms in eq.\ 1 are a single component, the width of this component (in $\alpha$) allows both of these terms to be
simultaneously non-zero in that equation.  In consequence, as time progresses, the second term in eq.\ 1 and then the third will
gain amplitude, but the third cannot do that until after a time $T$ that corresponds to the time it takes for the mechanical
device to complete its task.

Since we arranged to have the first component decrease for a time equal to the half-life of a single emission, its square modulus
will stabilize to a constant value of 0.5 at that time, assuming that eq.\ 1 is normalized.  After that, no new current will flow
into the second component, so the amplitude of $d_1M(\alpha, t)i_0$ will fall back to zero as the pulse $M(\alpha, t)$ runs out
along $\alpha$.  When $M(\alpha, t)$ finally goes to zero, the last term $d_1M(\alpha_f, t)i_1$ will reach its maximum value.  In
the end, the first component and the last term (representing the final form of the second component) will survive, each with a
square modulus equal to 0.5.

\section*{Add an Observer}
Before inflicting this apparatus on a cat, we will see how the oRules work when an outside observer witnesses the apparatus in
operation -- i.e., after the primary interaction has begun.  In that case, eq.\ 1 becomes
\begin{equation}
\Phi(t \ge t_0)= [d_0M(\alpha_0, t)i_0 + d_1M(\alpha,t)i_0 \rightarrow d_1M(\alpha_f, t)i_1]\otimes X
\end{equation}
where $X$ is the unknown brain state of the observer prior to the physiological interaction. Let the observer look at the
detector at some time $t_{look}$, giving
\begin{eqnarray}
\Phi(t \ge t_{look} > t_0) &=& d_0M(\alpha_0, t)i_0\otimes X \rightarrow d_0M'(\alpha_0, t)I_0B^b  \nonumber \\
&+& d_1M(\alpha, t)i_0\otimes X \rightarrow d_1M'(\alpha, t)I_0B^b  \nonumber \\
&\rightarrow& d_1M(\alpha_f, t)i_1\otimes X \rightarrow d_1M'(\alpha_f, t)I_1B^b \nonumber
\end{eqnarray}
where the physiological process represented by the arrows is a continuous and classical progression leading from independence to
entanglement.  The arrows carry $i$ into $I$ and $\otimes X$  into $B^b$.  The indicator $I$ includes the bare device $i$
\emph{plus} the low-level physiology of the observer. The brain state $B$ is only the higher level processes of the brain that
directly involve consciousness (i.e., not image processing). The state $B^b$ is called a \emph{brink state} because it is an
inactive state (i.e.,
 not yet conscious or ready) that is on the brink of becoming active.  This progression is classical to the
extent that it ignores the quantum uncertainties that are internal to the apparatus and the physiology of the observer\footnote{The
``decision" of the observer to look at the detector is assumed to be deterministically internal in an \emph{ontological} model
like this one (see ref.\ 1).  In this respect, any ontological model is like classical physics.}.  

During this process the observer will be unable to distinguish between the two detector states $d_0$ and $d_1$, which is why his
brain is called inactive in the above equation.  However, at some moment $t_{ob}$ (i.e., the moment of observation) the
observer will resolve the difference between these states, and when that happens a continuous `classical' evolution will no
longer be possible.  Let this happen with the appearance of the brink state $B^b$.  At this point the solution will branch
``quantum mechanically" into additional components, with \mbox{oRule (2)} requiring the introduction of ready brain states
$\underline{B}_0$ and
$\underline{B}_1$.  
\begin{eqnarray}
&&\Phi(t \ge t_{ob} >  t_{look} > t_0) =\\
 &=& d_0M'(\alpha_0, t)I_0B^b + d_1M'(\alpha, t)I_0B^b \rightarrow d_1M'(\alpha_f, t)I_1B^b\nonumber  \\
&+& d_0M''(\alpha_0, t)I_0\underline{B}_0 (+) d_1M''(\alpha, t)I_0\underline{B}_0 (\rightarrow) d_1M''(\alpha_f,
t)I_1\underline{B}_1 
\nonumber 
\end{eqnarray}
where $\underline{B}_0$ and $\underline{B}_1$ are ready brain states that interact with of $I_0$ and $I_1$ 
respectively\footnote{$\underline{B}_0$ is a ready state in the fifth/sixth term of eq.\ 3 because that component is
discontinuous with the initial component. $\underline{B}_0$  is a ready state in the fourth term because is
discontinuous with the fifth/sixth term as per oRule (2).  Each passage from $B^b$ to $B_0$ in this equation occurs at the same
time, so if the observer looks at the apparatus after the primary interaction has begun, both the fourth and the fifth/sixth terms
in eq.\ 3 will be simultaneously non-zero.}. Probability current in eq.\ 3 flows vertically into the second row from components in
the first row, beginning at the time $t_{ob}$.  This current represents the physiological interaction that follows the active
engagement of the observer.  Horizontal current will flow in the first row during the primary interaction, but not in the second
row because of oRule (4) which says that the ready brain state $\underline{B}_0$ cannot transmit current to the ready brain state
$\underline{B}_1$, or to itself.   Forbidden current flow is indicated by the parenthesis around the plus sign and the arrow in
the second row of eq.\ 3. 

\vspace{.4cm}

The first oRule requires that the time integrated current flowing into the second row of eq.\ 3 must equal 1.0.  So one of the
recipient terms must be eventually chosen.  

If the observation occurs after time $T$, then current will have gone into the last (sixth) term $d_1M''(\alpha_f,
t)I_1\underline{B}_1$ in eq.\ 3.  If that term happens to be stochastically chosen at a time $t_{sc6}$, then the ready state
$\underline{B}_1$ will become conscious, and following oRule (3) the state of the system will be
\begin{displaymath}
\Phi(t \ge t_{sc6} > t_{ob}) = d_1M(\alpha_f, t_{sc6})I_1B_1
\end{displaymath}
This will complete the interaction.  It corresponds to the observer coming on board when the mechanical device has already
finished its task.  The \mbox{un-under}-lined state $B_1$ is a brain state of the observer that is conscious of the indicator
$I_1$ (and possibly the detector $d_1$).

If the fifth term in eq.\ 3 is stochastically chosen at $t_{sc5}$, then 
\begin{displaymath}
\Phi(t \ge t_{sc5} > t_{ob}) = d_1M(\alpha, t)I_0B_0 \rightarrow d_1M(\alpha_f, t)I_1B_1
\end{displaymath}
beginning at time $t_{sc5}$ before the final time $t_f$.  So the final state is 
\begin{equation}
 \Phi(t \ge t_f > t_{sc5} > t_{ob}) =  d_1M(\alpha_f, t)I_1B_1 
\end{equation}

Another possibility will be that there will be a stochastic hit on the fourth term $d_0M''(\alpha_0, t)I_0\underline{B}_0$ in eq.\
3 at a time
$t_{sc4}$.  This reduces the state of the system to  
\begin{displaymath}
\Phi(t = t_{sc4} > t_{ob}) = d_0M(\alpha_0, t)I_0B_0
\end{displaymath}
which will continue to evolve under the primary interaction giving	
\begin{equation}
\Phi(t \ge t_{sc4} > t_{ob}) = d_0M(\alpha_0, t)I_0B_0 + d_1M'(\alpha_0, t)I_0\underline{B}_0
\end{equation}
where the second component is zero at $t_{sc4}$ and increases in time.  This component will not be succeeded by another value of 
$\alpha$ in the continuum that normally follows $d_1M'(\alpha_0, t)I_0\underline{B}_0$ because   oRule (4)  will not allow a
self-generating succession of ready brain states.  That is, no transition is allowed \emph{from} the second component in eq.\ 5
because it contains the ready brain state $\underline{B}_0$.  Consequently, the component $\alpha_0$ \emph{cannot be skipped
over} as the mechanical device begins its operation.  

The second component in eq.\ 5 may now be stochastically chosen at time $t_{sc42}$, such that $t_{sc42} > t_{sc4}$.  In that
case the system will again be reduced, giving
\begin{displaymath}
\Phi(t = t_{sc42} > t_{sc4} > t_{ob}) = d_0M(\alpha_0, t)I_0B_0
\end{displaymath}
From this point on, the observer will track the classical behavior of the mechanical   device as happened following eq.\ 4,
beginning in this case with $\alpha_0$ and ending with $\alpha_f$ in $d_1M(\alpha_f, t)I_1B_1$.

\vspace{.4cm}

	The primary (horizontal) current will cut off at $t_{1/2}$.  If that happens after the fourth term in eq.\ 3 has been
stochastically chosen at $t_{sc4}$, and before eq.\ 5 has run its course, it (eq. 5) will become time independent.
\begin{equation}
\Phi(t \ge t_{1/2} >t_{sc4} > t_{ob}) = d_0M(\alpha_0, t_{1/2})I_0B_0 + d_1M'(\alpha_0, t)I_0\underline{B}_0
\end{equation}
The existence of this superposition is like similar cases in the previous paper (ref.\ 1) where a component containing as ready
brain state no longer takes in probability current. We call that component a \emph{phantom} because it serves no further purpose
(see discussion in ref.\ 1).  This designation applies to the second component in eq.\ 6, so we choose to redefine the system by
dropping this component (again see ref.\ 1).  Equation 6 therefore corresponds to the observer (in the first component) finding
the detector in the state $d_0$ and the indicator in $I_0$ with the clock run out.  

	If the cut-off occurs after the fifth term in eq.\ 3 has been stochastic chosen, it (the fifth term) will continue to run its
classical course to the sixth component like eq.\ 4, ending with the observer being aware that the mechanical devise has
completed its task.

	And finally, if the cut-off at $t_{1/2}$ occurs before there has been a stochastic hit of any kind, then the second and third
terms of eq.\ 3 will go to zero as current is drained from them into the fifth and sixth terms.  This may give rise to a stochastic
hit on one of these terms, reducing the state to either the fifth term that subsequently evolves into the sixth term, or it will
reduce to the sixth term directly.  

Otherwise, without such a stochastic hit, the component that is the fifth and sixth terms will
become a phantom after receiving this current.  However, physiological (vertical) current will continue to flow after $t_{1/2}$
from the first component to the fourth term $d_0M''(\alpha_0, t)I_0\underline{B}_0$ in eq.\ 3.  This will lead to a hit on the
fourth term at some time
$t_{sc43}$ giving
\begin{equation}
\Phi(t \ge t_{sc43} >t_{1/2} > t_{ob}) = d_0M(\alpha_0, t_{sc43})I_0B_0 
\end{equation}
This again  corresponds to the observer finding the detector in detector state $d_0$ with the clock run out.  

Since the clock limiting the detector is set equal to the half-life of a single emission, there is a 50\% chance that the system
will terminate in the original state.  Otherwise, there is a 50\% chance that the mechanical device will go to the end.

\section*{Version 1 with no Outside Observer}
We now replace the indicator in eq.\ 1 with cat brain states.  The first of these is $C_0$, the brain state of the cat when it is
conscious of the variable $\alpha_0$.  This is the state of the cat before a stochastic hit.  It is now required  that
all lower physiological operations of the observer's brain are included in the mechanical device.  Before a stochastic choice
occurs, the system is given by 
\begin{equation}
\Phi(t \ge t_0) = d_0M(\alpha_0, t)C_0  + d_1M'(\alpha_0, t)\underline{C}_0
\end{equation}
where the second component is initially zero and increases in time.  The time dependence of $M$ and $M'$ affects only the
change of square modulus that results from the flow of current from the first to the second component.  The ready state in eq.\ 8
is not conscious; and in any case, it poses no  paradox of the kind generally associated with the cat.  The fourth rule
again insures that there cannot be a stochastic choice of alphas higher than $\alpha_0$, so again, $\alpha_0$ cannot be passed
over.  

If there is a stochastic hit at time $t_{sc}$, eq.\ 8 will become
\begin{equation}
\Phi(t_f \ge t \ge t_{sc} > t_0) = d_1M(\alpha_0, t_{sc})C_0 \rightarrow d_1M (\alpha, t)C_\alpha \rightarrow d_1M(\alpha_f, t_f)U
\end{equation}
where $U$ is the unconscious state of the cat, and $C_\alpha$ is the brain state of the cat when it is conscious of the variable
$\alpha$.  Again, $\alpha_0 \ge \alpha < \alpha_f$.  The two different brain states in this equation do not result in a paradox
because they occur at different times.  The three terms in eq.\ 9 represent one component at three different times.  Time
dependence in $M$ and $M'$ therefore refers to that evolution and not to a change of square modulus.  Also in this case,
$M(\alpha, t)$ is not a pulse of quantum mechanical uncertainties as it is in  eq.\ 1, for the process in eq.\ 9 is initiated by a
sharply defined stochastic hit.  So $\alpha$ has a sharply defined and classically determined value at any time $t$.  The arrows
in eq. 9 carry $t$ classically and continuously from $t_{sc}$ to $t_f$, as $\alpha_0$ goes to $\alpha_f$ and $C_0$ goes to $U$. 
The final state of the cat is therefore  
\begin{equation}
\Phi(t \ge  t_f) = d_1M(\alpha_f, t_f)U
\end{equation}

If, on the other hand, primary current stops flowing at the half-life time $t_{1/2}$ before a stochastic hit,  then the second
component in eq.\ 8 will stabilize in place giving the time independent equation
\begin{equation}
\Phi(t \ge t_{1/2} >t_0) = d_0M(\alpha_0, t_{1/2})C_0  + d_1M'(\alpha_0, t_{1/2})\underline{C}_0
\end{equation}
where both components have come to a square modulus equal to 0.5, assuming that the equation is initially normalized.  The cat in
eq.\ 11 remains conscious of $\alpha_0$, so  it has escaped being put to sleep.

There is a 0.5 probability that eq.\ 10 will be the final state, and a 0.5 probability that eq.\ 11 will be the final state.  This
 confirms our  expectations.

\section*{Version I with Outside Observer}
	Imagine that an outside observer looks in on the cat during these proceedings to see how it is doing.  This observer is initially
in the wings represented by $\otimes X$ as in eq.\ 2.  The physiological interaction applied to eq. 8 is then
\begin{eqnarray}
&&\Phi(t > t_{ob} \ge t_{look} >  t_0) =\\
 &=& d_0M(\alpha_0, t_{look})C_0\otimes X \rightarrow d_0M(\alpha_0, t_{ob})C_0B^b \rightarrow d_0M(\alpha_0, t)C_0B_0\nonumber 
\\ &+& d_1M'(\alpha_0, t_{look})\underline{C}_0\otimes X \rightarrow d_1M'(\alpha_0, t_{ob})\underline{C}_0B^b \rightarrow
d_1M'(\alpha_0, t)\underline{C}_0\underline{B}_0 
\nonumber 
\end{eqnarray}
where again $B^b$ is the brink state of the observer, and $B_0$ is the ready brain state of the observer when he is aware of the
cat being conscious of $\alpha_0$.  The first row of this equation is a single component that evolves continuously and classically
as represented by the arrow.  It carries $\otimes X$ into $B_0$ by a process that leads from independence to
entanglement\footnote{The realized state $B_0$ in the first row is the result of a continuous evolution arising from the
physiological interaction.   The ready state $\underline{B}_0$ in the second row is the result of the discontinuous evolution
arising from the primary interaction -- i.e., the sixth term is discontinuous with the initial state.}. The mechanical device
associated with
$B^b$ after time $t_{ob}$ now includes the lower level physiology of both the cat and the outside observer.  

The primary interaction is still active during this time, and this gives rise to a vertical current going from the first to the
second row in eq.\ 12.  The second row is therefore a continuum of terms that are created parallel to the first row at each
moment of time.  So after $t_{ob}$, vertical current flows only into the final component in the second row of eq.\ 12. 
Terms prior the last one no longer have current flowing into them from above, and since there is no horizontal current among
these ready states, they become phantoms as soon as they are created.  Therefore, when the interaction is complete eq.\ 12
is 
\begin{equation}
\Phi(t > t_{ob}) = d_0M(\alpha_0, t)C_0B_0  + d_1M'(\alpha_0, t)\underline{C}_0\underline{B}_0
\end{equation}
where $d_1M'(\alpha_0, t)\underline{C}_0\underline{B}_0$ is initially zero  and increases in time.  This equation is
identical with eq.\ 8 with the addition of the observer.  From this point on the options work out the same as they do in the
previous section, except for the addition of the observer. 

	If there is a stochastic hit between times $t_{look}$ and $ t_{ob}$ in eq.\ 12, then the corresponding ready state in the
second row will be stochastically chosen, and the subsequent continuous process will lead to the final state $d_1M'(\alpha_f,
t_f)UB_U$, where $B_U$ is the brain state of the observer when he is aware of the unconscious cat.

\section*{Version II with no Outside Observer}
	In the second version of the Schr\"{o}dinger cat experiment, the cat is initially unconscious, and is awakened by an alarm that is
set off by the capture of a radioactive decay.  The mechanical device $M(\alpha, t)$ is now an alarm clock, where $\alpha$ 
represents the clock's successive stages -- from its initial response (to radioactivity) to the ring.  As before, the alarm will
go off only 50\% of the time.  The equation of state is
\begin{equation}
\Phi(t \ge t_0) = d_0M(\alpha_0, t)U  + d_1M(\alpha, t)U \rightarrow d_1M(\alpha_f, t)\underline{C}
\end{equation}
where $U$ is the initial unconscious state of the cat, $\underline{C}$ is the cat's final (and still unconscious) ready brain
state, and the second and third terms (of the second component) are initially equal to zero and increase in time.  Again, there
may be a time delay
$T$ before the third term containing the ready brain state of the cat can accumulate value after $t_0$.  We assume eq.\ 14 to be
normalized.  

When current does flow into the third term, it might be stochastically chosen at time $t_{sc}$.  If that happens, 
the system will become
\begin{equation}
\Phi(t \ge t_{sc} > t_f >  t_0) = d_1M(\alpha_f, t_{sc})C
\end{equation}
This will terminate the interaction.  It corresponds to the cat finding himself aroused by the alarm 50\% of the time. 

Only the third term in eq.\ 14 contains a ready brain state, so only it can be stochastically chosen in a way that leads to an
oRule (3) reduction.  If there is no stochastic choice, then the square modulus of the first component of eq.\ 14 will fall to a
value of 0.5.  The second term will initially rise to some positive value and fall again to zero, and the third term will
rise to a square modulus of 0.5.  In the final state of the system, the square moduli of the first  and  third terms
will be equal to 0.5, and the second  will be zero.  Therefore, at and after $t_f$ when the alarm mechanism has run its
course, the system will end its evolution with two time independent components
\begin{displaymath}
\Phi(t \ge t_f) = D_0M(\alpha_0)U +  D_1M(\alpha_f)\underline{C}
\end{displaymath}
which will appear 50\% of the time.  The cat is not conscious in either one of them.  Redefining the system by dropping the second
component gives
\begin{displaymath}
\Phi(t \ge t_f) = D_0M(\alpha_0)U 
\end{displaymath}

\section*{Version II with Outside Observer}
If the outside observer interacts with the cat \& apparatus after there has been a stochastic choice leading to eq.\ 15, then
following a separate physiological interaction at $t_{ob}$, the conscious observer will be on board with the conscious cat.  The
two of them will experience an amended version of eq.\ 15 given by
\begin{equation}
\Phi(t \ge t_{ob} > t_{sc} > t_f > t_0) = D_1M(\alpha_f)CB_{fC}
\end{equation}
where $B_{fC}$ is the observer's state of awareness of the variable $\alpha_f$ \emph{and} the conscious cat C.  

Now imagine that the outside observer enters the picture before the stochastic choice that leads to eq.\ 15, assuming that the
primary interaction has already begun.  Equation 14 will be
\begin{eqnarray}
\Phi(t &\ge& t_{ob}) =\\
 &=& d_0M(\alpha_0, t)UB^b + d_1M(\alpha, t)UB^b \rightarrow d_1M(\alpha_f, t)\underline{C}B^b\nonumber 
\\ &+& d_0M'(\alpha_0, t)U\underline{B}_{0 U} (+) d_1M'(\alpha, t)U\underline{B}_{\alpha U}  
\nonumber 
\end{eqnarray}
where $\underline{B}_{0U}$ (or $\underline{B}_{\alpha U}$) is the observer's ready state of awareness of the variable $\alpha_0$
(or $\alpha$) and the unconscious cat. The fourth oRule blocks the appearance of a sixth term.  The ready components in the second
row are equal to zero at $t_{ob}$ and increase in time\footnote{Again, $\underline{B}_{\alpha U}$ is a ready state because it is
discontinuous with the initial state, and
$\underline{B}_{0U}$ is a ready state because it is discontinuous with $\underline{B}_{\alpha U}$.}.  The appearance of the
observer leads to a stochastic hit on one of these components.  

If $d_0M'(\alpha_0, t)U\underline{B}_{0U}$ is chosen at time $t_{sc4}$, then eq.\ 17 reduces to
\begin{equation}
\Phi(t \ge t_{sc4} >  t_{ob}) = d_0M(\alpha_0, t)UB_{0U} + d_1M'(\alpha_0, t)U\underline{B}_{0U}
\end{equation}
where the second component is zero at $t_{sc4}$ and increases in time. This component cannot advance further than $\alpha_0$
because of oRule (4).  Another stochastic hit at $t_{sc42}$ will give
\begin{displaymath}
\Phi(t \ge t_{sc42} > t_{sc4} > t_{ob}) = d_1M(\alpha_0, t)UB_{0U} \rightarrow  d_1M(\alpha_f, t)CB_{fC}
\end{displaymath}
resulting in
\begin{equation}
\Phi(t \ge t_f > t_{sc42} > t_{sc4} >  t_{ob}) =  d_1M(\alpha_f, t_{sc42})CB_{fC}
\end{equation}
If $d_1M'(\alpha, t)U\underline{B}_{\alpha U}$ in eq.\ 17 is chosen at time $t_{sc5}$, then it reduces to
\begin{displaymath}
\Phi(t \ge t_{sc5} > t_{ob}) = d_1M(\alpha, t)UB_{\alpha U} \rightarrow  d_1M(\alpha_f, t)CB_{fC}
\end{displaymath}
resulting in
\begin{equation}
\Phi(t \ge t_f > t_{sc5} >  t_{ob}) =  d_1M(\alpha_f, t_{sc5})CB_{fC}
\end{equation}

If there is a stochastic hit on the third term in eq.\ 17 at $t_{sc3}$,  it will evolve continuously to become 
\begin{equation} 
\Phi(t \ge t_{sc3} > t_f > t_{ob}) = d_1M(a_f, t_{sc3})CB_{fC} 
\end{equation}
In either case, the final state in eqs. 19, 20 or 21 is the same as eq.\ 15 except that the external observer in now on board.

\section*{Version II with a Natural Wake-Up}
	Even if the alarm does not go off, the cat will wake up naturally by virtue of its own internal alarm clock.  The internal alarm
can be represented by a classical  mechanical device that operates at the same time as the external alarm.  The interaction is
assumed to run parallel to eq.\ 14, and is given by  
\begin{equation}
\Phi(t \ge   t_0) =  N(t_0)U \rightarrow N(t)U \rightarrow N(t_{ff})C
\end{equation}
where $N(t)$ is the internal mechanism and $t_{ff}$ is the final time of its internal development.  Taking the product of eqs.\ 14
and 22 at $t_0$ gives
\begin{displaymath}
\Phi(t = t_0) = d_0M(\alpha_0, t)N(t)U 
\end{displaymath}
After which the state becomes the subsequent product
\begin{eqnarray}
\Phi(t \ge t_0) &=& [d_0M(\alpha_0, t)U + d_1M(\alpha, t)U \rightarrow d_1M(\alpha_f, t)\underline{C}] \nonumber\\
&\times& [N(t_0)U \rightarrow N(t)U \rightarrow N(t_{ff})C] \nonumber 
\end{eqnarray}
where the cross product suggests a conflict between $C$ and $U$ states.  To resolve this, we follow two possible scenarios.  The
first assumes that the stochastic choice and  external decay occurs before the internal decay, and the second assumes that the
internal decay occurs before the stochastic choice and  external decay.  The first of these gives
\begin{eqnarray}
\Phi(t \ge t_0) &=& [d_0M(\alpha_0, t)U + d_1M(\alpha, t)U \rightarrow d_1M(\alpha_f, t)\underline{C}] \nonumber\\
&\times& [N(t_0) \rightarrow N(t)] \nonumber 
\end{eqnarray}
resulting in
\begin{eqnarray}
\Phi(t \ge t_{ff} > t_f > t_{sc} > t_0) &=& d_1M(\alpha_f, t_f)[N(t_0) \rightarrow N(t) \rightarrow N(t_{ff})]C
\nonumber\\ &=& d_1M(\alpha_f, t_f)N(t_{ff})C  
\end{eqnarray}

The second scenario is
\begin{displaymath}
\Phi(t \ge t_0) = [N(t_0)U \rightarrow N(t)U \rightarrow N(t_{ff})C][d_0M(\alpha_0, t) + d_1M(\alpha, t)]
\end{displaymath}
After $t_{ff}$ this becomes
\begin{displaymath}
\Phi(t \ge t_{ff} > t_0) = N(t_{ff})][d_0M(\alpha_0, t) + d_1M(\alpha, t) \rightarrow d_1M(\alpha_f, t)]C
\end{displaymath}
And after $t_f$ 
\begin{equation}
\Phi(t \ge t_f > t_{ff} >  t_0) =  d_1N(t_{ff})M(\alpha_f, t_f)C
\end{equation}
Both of these scenarios lead to the same conscious state in eqs.\ 22 and 23.

\end{document}